\documentclass{article}

\usepackage{macros_jgrange}

\newcommand{\simindex}{\alpha}
\newcommand{\simindexbis}{\beta}
\NewDocumentCommand{\structcounter}{O{\simindexbis}}{\structgraphe_{#1}}
\NewDocumentCommand{\structcounterbis}{O{\simindexbis}}{\structgraphebis_{#1}}

\NewDocumentCommand{\deccounter}{O{\simindexbis}}{\mathcal D(\structgraphe_{#1})}
\NewDocumentCommand{\deccounterbis}{O{\simindexbis}}{\mathcal D(\structgraphebis_{#1})}

\newcommand{\counterfam}{(\structcounter)_{\simindexbis\in\N}}
\newcommand{\counterfambis}{(\structcounterbis)_{\simindexbis\in\N}}

\title{On the nonexistence of \FO-continuous path and tree-decompositions}

\author{Julien Grange}
\date{}

\begin{document}

\maketitle

\begin{abstract}

  Bojanczyk and Pilipczuk showed in their celebrated article \emph{Definability equals recognizability for graphs of bounded treewidth} (LICS 2016) that monadic second-order logic can define tree-decompositions in graphs of bounded treewidth. This raises the question whether such decompositions can already be defined in first-order logic (\FO).

  We start by introducing the notion of tree-decompositions of bounded span, which restricts the diameter of the subtree consisting of the bags containing a same node of the structure. Having a bounded span is a natural property of tree-decompositions when dealing with \FO, since equality of nodes cannot in general be recovered in \FO when it doesn't hold. In particular, it encompasses the notion of domino tree-decompositions.

  We show that path-decompositions of bounded span are not \FO-continuous, in the sense that there exist arbitrarily \FO-similar graphs of bounded pathwidth which do not possess \FO-similar path-decompositions of bounded span. Then, we show that tree-decompositions of bounded span are not \FO-continuous either.
\end{abstract}

\section{Introduction}

The notion of treewidth was introduced in 1986 by Robertson and Seymour~\cite{DBLP:journals/jal/RobertsonS86} to measure how far a given graph is from being a tree. Roughly speaking, a graph $\structgraphe$ has treewidth $k$ if there exists a tree, called a tree-decomposition for $\structgraphe$, whose nodes are bags of at most $k+1$ vertices of $\structgraphe$, and whose structure mimics that of $\structgraphe$.

This gives a measure of sparsity for classes of graphs: a class $\classe$ has bounded treewidth if there exists a constant that bounds the treewidth of every graph belonging to $\classe$.

\bigskip

Classes of bounded treewidth have been widely studied, since they enjoy many nice properties; not least of all, the model checking of a \CMSO (monadic second-order logic with counting) sentence on graphs of bounded treewidth is decidable in linear time~\cite{DBLP:journals/iandc/Courcelle90}.

In 2016, Bojanczyk and Pilipczuk~\cite{DBLP:conf/lics/BojanczykP16} solved the long-lasting Courcelle conjecture: over classes of graphs of bounded treewidth, is definability in \CMSO equivalent to recognizability? This is reminiscent of the famous B\"uchi~\cite{buchi1960weak} equivalence over words and trees between definability in \MSO and recognizability by a finite automaton. In the context of graphs, this equivalence is significantly harder to get, due to the lack of structure. The technique used by Bojanczyk and Pilipczuk, on top of allowing to prove the conjecture, is very profound on its own: they show that it is possible, in \MSO, to define (through an \MSO-transduction) tree-decompositions inside graphs of bounded treewidth.

This immediately yields the question whether such a transformation is already definable in \FO (first-order logic).  We show in this article that this is not the case, at least when we restrict our study to those decompositions in which all the bags containing a single node of the graph are close together in the decomposition. We call this property \emph{having bounded span}. Having bounded span is a very natural and weak requirement:

\begin{itemize}
\item very natural, in particular in the context of \FO, since without it recovering the equality relation between nodes in the original structure is not possible.
\item weak in that it generalizes the notion of domino treewidth~\cite{DBLP:journals/jal/BodlaenderE97}, in which a single node can appear in at most two bags. Note that contrary to domino treewidth, having a tree-decomposition of bounded span does not bound the degree of a graph.
\end{itemize}

More precisely, we give a negative answer to two following questions, which implies in a strong sense that path and tree-decompositions of bounded span are not \FO-definable. Not only does this entail that there is no hope to find \FO-interpretations which define decompositions in graphs, but it also excludes the possibility of finding decompositions in a non-uniform way.

\newcommand{\width}{\texttt{width}}
\newcommand{\spa}{\texttt{span}}

\begin{question}[Are path-decompositions of bounded span \FO-continuous?]
  \label{qu:pw}

    Are there functions $\width,\spa:\N^2\to\N$ and $f:\N^3\to\N$, such that, given
  
  \begin{itemize}
  \item a width $w\in\N$, and a span $s\in\N$,
  \item some similarity index $\simindex\in\N$, and
  \item two graphs (or more generally, two structures) $\structgraphe$ and $\structgraphebis$ that admit path-decompositions of width at most $w$ and of span at most $s$, such that $\structcounter[]$ and $\structcounterbis[]$ agree on all \FO-sentences of quantifier rank at most $f(w,s,\simindex)$,
  \end{itemize}

  there exist two respective path-decompositions $\deccounter[]$ and $\deccounterbis[]$ of $\structcounter[]$ and $\structcounterbis[]$, which have width at most $\width(w,s)$ and span at most $\spa(w,s)$, such that $\deccounter[]$ and $\deccounterbis[]$ agree on all \FO-sentences of quantifier rank at most $\simindex$.

\end{question}

To rephrase in simpler terms, a positive answer to Question~\ref{qu:pw} would mean that provided that two graphs of bounded pathwidth are similar enough in the eyes of \FO, then they admit path-decompositions of bounded span that are similar wrt. \FO. This is why we assign the term of \emph{\FO-continuity} to this notion. Note that in this process, we allow the width and span of the decompositions to be as large as one wishes.

Let's now state a similar question in the case of tree-decompositions:

\begin{question}[Are tree-decompositions of bounded span \FO-continuous?]
  \label{qu:tw}
  Are there functions $\width,\spa:\N^2\to\N$ and $f:\N^3\to\N$, such that, given
  
  \begin{itemize}
  \item a width $w\in\N$, and a span $s\in\N$,
  \item some similarity index $\simindex\in\N$, and
  \item two graphs (or more generally, two structures) $\structgraphe$ and $\structgraphebis$ that admit tree-decompositions of width at most $w$ and of span at most $s$, such that $\structcounter[]$ and $\structcounterbis[]$ agree on all \FO-sentences of quantifier rank at most $f(w,s,\simindex)$,
  \end{itemize}

  there exist two respective tree-decompositions $\deccounter[]$ and $\deccounterbis[]$ of $\structcounter[]$ and $\structcounterbis[]$, which have width at most $\width(w,s)$ and span at most $\spa(w,s)$, such that $\deccounter[]$ and $\deccounterbis[]$ agree on all \FO-sentences of quantifier rank at most $\simindex$.

\end{question}

Note that a positive answer to Question~\ref{qu:tw} does not immediately yield a positive answer to Question~\ref{qu:pw}, since it would only guarantee the existence of fitting tree-decompositions, but would say nothing about path-decompositions.

In this paper, we exhibit families of structures (more precisely, families of colored graphs) which answer both questions in the negative.

Since it can easily be seen that ``having pathwidth (resp. treewidth) at most $k$'' are not \FO-definable properties, the fact that we answer these questions by the negative is not necessarily surprising. However, these negative results do not \textit{a priori} follow from the first remark, and showing that some arbitrarily similar graphs admit no similar decompositions proves to be much harder than proving the former statement.

\bigskip

As a consequence, the negative answer to Question~\ref{qu:tw} excludes the possibility to lift results involving \FO from trees to structures of bounded treewidth through their decompositions. For instance, when trying to extend the result from \cite{DBLP:journals/jsyml/BenediktS09} that \[\oifo=\FO\text{ on trees}\]
to classes of bounded treewidth, the most natural way would be to prove that, for various degrees of \FO-similarity,
\begin{enumerate}[(i)]
\item\label{enu:down} given two graphs of fixed treewidth that are \FO-similar, it is possible to find \FO-similar tree-decompositions (possibly not of optimal width), and
\item\label{enu:up} there exists an \FO-interpretation which defines a graph of fixed treewidth in one of its tree-decompositions.
\end{enumerate}

For the way up (\ref{enu:up}), from decompositions back to the graphs, one needs to be able to identify nodes belonging to different bags of the decomposition, and which correspond to the same node in the graph. As we have seen, this corresponds to controlling the span of the decomposition.

However, as soon as we ask for the decompositions to have a bounded span, a negative answer to Question~\ref{qu:tw} asserts that there is no way down (\ref{enu:down}), from structures to their decompositions.

Hence, in order to lift results involving \FO from trees to classes of bounded treewidth, one has to adopt a strategy that doesn't involve tree-decompositions, e.g. as in~\cite{grange_et_al:LIPIcs:2020:11666}. One could argue that in view of this, the treewidth is not a good measure of sparsity when working on \FO, as its definition cannot be leveraged in this context.

\paragraph{Overview of the article}

We introduce in Section~\ref{sec:prelim} a logical framework to speak of path and tree-decompositions, and we define there the span of a decomposition.

In Section~\ref{sec:pw}, we exhibit a counter-example that yields a negative answer to Question~\ref{qu:pw}. The general method is to assume, towards a contradiction, the existence of \FO-similar path-decompositions of bounded span to specific \FO-similar colored graphs. These graphs are constructed in such a way that, did they admit \FO-similar path-decompositions, these decompositions would have a small diameter; and in fact, too small for every node of the initial structures to appear in them.

To bring as well a negative answer to Question~\ref{qu:tw}, we use again some of the ideas from Section~\ref{sec:pw} in Section~\ref{sec:tw}, although the manipulation requires a bit of extra care. The additional difficulties encountered when going from pathwidth to treewidth are essentially due to the fact that the degree of tree-decompositions is not bounded, which makes it harder to come to a contradiction when supposing the existence of \FO-similar tree-decompositions. Similarly as in Section~\ref{sec:pw}, we get that any \FO-similar tree-decompositions of our colored graphs of choice must have a small diameter. However, since the degree of the decomposition is not restricted, we must first carefully trim these decompositions in order to derive a contradiction.

\paragraph{Related works}

The notion of decomposition of bounded span is a relaxation of the notion of domino decomposition, as introduced in~\cite{DBLP:journals/jgt/DingO95} and~\cite{DBLP:journals/jal/BodlaenderE97}. More precisely, we do not restrict the number of bags containing a single node, but only the distance between them.

As for our result on treewidth, it can be seen as a negative counterpoint in \FO to the positive result, for \MSO, of Bojanczyk and Pilipczuk~\cite{DBLP:conf/lics/BojanczykP16}. Indeed, the negative answer to Question~\ref{qu:tw} given in Section~\ref{sec:tw} entails, in particular, that tree-decompositions (at least when their span is bounded) are not \FO-interpretable.

\section{Preliminaries}
\label{sec:prelim}

\subsection{Logic}

We use the standard definitions and notations for logics and structures (see, e.g., \cite{DBLP:books/sp/Libkin04}).

Given two $\vocab$-structures $\struct$ and $\structbis$, we say that \textbf{$\struct$ is $\FO$-similar to $\structbis$ at depth $k$}, and we note $\struct\foeq{k}\structbis$, if $\struct$ and $\structbis$ agree on every $\FO$-sentence $\formule$ of quantifier rank at most $k$; that is, if $\struct\models\formule$ iff $\structbis\models\formule$ for every such $\formule$.

We write $\struct\simeq\structbis$ if $\struct$ and $\structbis$ are isomorphic. We denote the distance between the elements $x$ and $y$ in $\struct$ by $\dist{\struct}{x}{y}$: this is the distance in the Gaifman graph of $\struct$, where two elements are at distance $1$ iff they appear in the same tuple of some relation.

The classical way to define new structures from existing ones in \FO is through \FO-interpretations. An \FO-interpretation $\inter$ from a vocabulary $\vocab$ to a vocabulary $\vocab'$ is a tuple of \FO-formul\ae\ on $\vocab$:

\begin{itemize}
\item first, a formula $\phi$ with $r$ free variables, which defines the domain of the new structure $\inter(\struct)$ as a subset of some cartesian power $\structdom^r$ of the domain of the initial structure $\struct$,
\item then, for every relation symbol $R\in\vocabbis$ of arity $a$, a formula $\varphi_R$ with $a\cdot r$ free variables, which gives the interpretation in $\inter(\struct)$ of the relation $R$.
\end{itemize}
The integer $r$ is the called the arity of $\inter$, while the depth of $\inter$ is the maximum among the quantifier ranks of $\phi$ and all the $\varphi_R$.

\FO-interpretations are the prime way to define uniform \FO-continuous transformation of structures, as \FO-similarity of the source structures yields \FO-similarity of the interpreted structures:

\begin{thm}
  \label{prop:interpretations_are_continuous}
  Let $\inter$ be an \FO-interpretation from $\vocab$ to $\vocabbis$, of arity $r$ and depth $d$.

  For every $\vocab$-structures $\struct,\structbis$ and for every $k\in\N$,
  \[\struct\foeq{rk+d}\structbis\quad\rightarrow\quad\inter(\struct)\foeq{k} \inter(\structbis)\,.\]
\end{thm}

\subsection{Treewidth and pathwidth}

Let us briefly recall the classical definitions of treewidth and pathwidth, before we translate them in logical terms in Section~\ref{sec:logical_persp_decomp}.

We consider graphs (and trees) as logical structures over the vocabulary $\vocabgraphe*$.

Let $\structgraphe*$ be a graph. A tree-decomposition of $\structgraphe$ is a tree $\structarbre*$ together with a function $\bag$ from $\arbredom$ to the power set of $\graphedom$, such that

\begin{itemize}
\item every element $x\in\graphedom$ appears in some bag: $\forall x\in\graphedom,\exists t\in\arbredom, x\in\bag(t)$,
\item for every edge $(x,y)$ of $\structgraphe$, $\exists t\in\arbredom. x\in\bag(t)\land y\in\bag(t)$,
\item for every element $x\in\graphedom$, $\{t\in\arbredom:x\in\bag(t)\}$ is connected in $\structarbre$.
\end{itemize}

It will be convenient to assume that a bag is always non-empty.

A tree-decomposition $\treedec$ is a path-decomposition if $\structarbre$ is a directed path.

The width of the tree or path-decomposition $\treedec$ is \[\max\{|\bag(t)|:t\in\arbredom\}-1\,.\]

\subsection{Logical perspective on decompositions}
\label{sec:logical_persp_decomp}

To give a meaning to the notion of \FO-similarity between two decompositions, we need to fix a logical framework for these decompositions. In the process, we extend these definitions to the context of purely relational structures, not only graphs. We omit the constant symbols for simplicity's sake, but one could consider vocabulary with constant with an extra bit of care.

Intuitively, a tree-decomposition is a colored tree, where the color of a node describes the content of the corresponding bag (i.e. the isomorphism class of the substructure induced by the bag) as well as how to glue that bag to the bags of the parent and the children of the node.

\bigskip

\NewDocumentCommand{\vocabbag}{O{k}}{\Lambda_{#1}}
\NewDocumentCommand{\isoclassbag}{O{k}}{\sigma_{#1}}
\NewDocumentCommand{\vocabdec}{O{k}}{\vocab_{#1}}
\NewDocumentCommand{\bagdom}{O{}}{B_{#1}}
\NewDocumentCommand{\structbag}{O{}}{\mathcal{\bagdom}_{#1}}

Let $\vocab$ be a relation vocabulary without constant symbols. For $k\in\N$, let's consider the vocabulary $\vocabbag:=\vocab\cup\{I_0,\cdots,I_k,O_0,\cdots,O_k\}$ where all the $I_i,O_i$ are fresh unary relation symbols. Bags of a decomposition of width $k$ are seen as $\vocabbag$-structures, where all the relations in $\vocab$ are inherited from the underlying structure, and the $I_i$ (resp. $O_i$) contain the elements that also appear in the parent's bag (resp. some child's bag).

More precisely, we say that a $\vocabbag$-structure $\structbag$ is a \textbf{$k$-bag} if it satisfies the following conditions:

\begin{itemize}
\item $|\bagdom|\leq k+1$
\item for every $0\leq i\leq k$, $|\interrel{I_i}{\structbag}|\leq 1$ and $|\interrel{O_i}{\structbag}|\leq 1$
\item for every $0\leq i<j\leq k$, $\interrel{I_i}{\structbag}\cap \interrel{I_j}{\structbag}=\interrel{O_i}{\structbag}\cap \interrel{O_j}{\structbag}=\emptyset$.
\end{itemize}

Let $\isoclassbag$ be the (finite) set of isomorphism classes of $k$-bags, and let \[\vocabdec:=\{P_c:c\in\isoclassbag\}\cup\{S\}\] be composed of new relation symbols, where the $(P_c)_{c\in\isoclassbag}$ are unary and $S$ is binary. For $c\in\isoclassbag$, we will write $c\models\formule$ if any (or equivalently, every) $\structbag\in c$ satisfies $\formule$.

We say that a $\vocabdec$-structure $\structarbre$ is a \textbf{valid tree-decomposition of width $k$} if

\begin{itemize}
\item its restriction to $\{S\}$ is an unranked tree, meaning that an element may have an arbitrary number of children, and makes no difference among them,
\item the sets $\interrel{P_c}{\structarbre}$, for $c\in\isoclassbag$, partition its domain $\arbredom$; in other words, every node is colored by exactly one color,
\item for every $c\in\isoclassbag$, for every $t\in\arbredom$ such that $\structarbre\models P_c(t)$ and for every $i\in\{0,\cdots,k\}$,

  \begin{tabular}{lll}
    \multirow{2}{*}{$c\models\exists x, O_i(x)$}&\multirow{2}{*}{iff}&there is a child $u$ of $t$ in $\structarbre$ such that $\structarbre\models P_d(u)$\\
    &&with some $d\in\isoclassbag$ such that $d\models\exists y,I_i(y)$,\\
  \end{tabular}
\item if $\structarbre\models P_c(r)$ where $r$ is the root of $\structarbre$, then $c\models\bigwedge_{0\leq i\leq k}\neg\exists x, I_i(x)\,.$
\end{itemize}

Basically, these conditions amount to saying that an \emph{output} element with the label $O_i$ is always interfaced with a corresponding \emph{input} element with the label $I_i$ in some child's bag, and conversely. The intent is to merge those elements in the construction to come.

Let $\TD{k}$ be the class of all valid tree-decompositions of width $k$, and $\PD{k}$ the subclass of $\TD{k}$ containing all the tree-decompositions whose restriction to $\{S\}$ is a directed path (i.e. a tree with a single branch).

\bigskip

Every $\structarbre\in\TD{k}$ generates a $\vocab$-structure $\ext$ in the natural way: we take the disjoint union of the $\vocab$-structures which correspond to the colors of all the elements of $\structarbre$, and we identify in this union an element colored with $I_i$ with the element colored with the corresponding $O_i$ in the parent's bag.


More precisely, $\ext$ is constructed as follows. For every $c\in\isoclassbag$ and every element $t\in\arbredom$ belonging to $\interrel{P_c}{\structarbre}$, we consider a $k$-bag $\structbag[t]$ belonging to $c$, such that the $(\structbag[t])_{t\in\arbredom}$ have pairwise disjoint domains. Let $\structdom:=\bigcup_{t\in\arbredom}\bagdom[t]$. We define the interface relation $I$ on $\structdom$, such that $I(x,y)$ holds iff $x\in\bagdom[t]$ and $y\in\bagdom[u]$ for some $t,u\in\arbredom$, such that

\begin{itemize}
\item $\structarbre\models S(t,u)$, i.e. $t$ is the parent of $u$ in $\structarbre$
\item $\structbag[t]\models O_i(x)$ for some $0\leq i\leq k$
\item $\structbag[u]\models I_i(x)$ for the same $i\,.$
\end{itemize}

Now, let $\sim$ be the most coarse-grained equivalence relation extending $I$. $\ext$ is defined as follows:

\begin{itemize}
\item its domain is $\quotient{\structdom}{\sim}$
\item for every $r$-ary relation symbol $\relsymb\in\vocab$,
  
  for every equivalence classes $\bar x_1,\cdots,\bar x_r$ for $\sim$, 
  
  $\ext\models\relsymb(\bar x_1,\cdots,\bar x_r)$ iff there exist $t\in\arbredom$, and $x_1,\cdots,x_n\in\bagdom[t]$ such that $x_i\in\bar x_i$ ($1\leq i\leq r$) and $\structbag[t]\models\relsymb(x_1,\cdots,x_r)\,.$
\end{itemize}

Roughly speaking, a tuple belongs to the interpretation of $\relsymb$ in $\ext$ iff it appears in some bag of $\mathcal T$.

\begin{figure}[!ht]
  \centering
  \begin{tikzpicture}
    \coordinate (a1) at (0,7);
    \coordinate (a2) at (2,7);
    \coordinate (b1) at (0,4);
    \coordinate (b2) at (2,4);
    \coordinate (c1) at (6,7);
    \coordinate (c2) at (8,7);
    \coordinate (d1) at (6,4);
    \coordinate (d2) at (8,4);

    \node at (a1) {
      \begin{tikzpicture}[scale=1]
        \draw (0,0) node{\color{red}{$\blacksquare$} \color{black} :};
      \end{tikzpicture}
    };

    \node at (a2) {
      \begin{tikzpicture}[scale=.3]
        \coordinate (0) at (1,1);
        \coordinate (1) at (0,0);
        \coordinate (2) at (2,0);
        
        \node[below] at (1) {\small $O_0$};
        \node[below] at (2) {\small $O_1$};
        \draw (0) -- (1) -- (2) -- (0);
        \draw[mygreen] (0) node{\small $\bullet$};
        \draw[red] (1) node{\small $\bullet$};
        \draw[myyellow] (2) node{\small $\bullet$};
      \end{tikzpicture}
    };

    \node at (b1) {
      \begin{tikzpicture}[scale=1]
        \draw (0,0) node{\color{myyellow}{$\blacksquare$} \color{black} :};
      \end{tikzpicture}
    };

    \node at (b2) {
      \begin{tikzpicture}[scale=0.3]
        \coordinate (0) at (1,-1);
        \coordinate (1) at (0,0);
        \coordinate (2) at (2,0);
        
        \node[below] at (0) {\small $O_0$};
        \node[above] at (1) {\small $I_0$};
        \node[above] at (2) {\small $I_1$};
        \node[below right] at (2) {\small $O_2$};
        \draw (0) -- (1) -- (2);
        \draw[blue] (0) node{\small $\bullet$};
        \draw (1) node{\small $\bullet$};
        \draw[myyellow] (2) node{\small $\bullet$};
      \end{tikzpicture}
    };

    \node at (c1) {
      \begin{tikzpicture}[scale=1]
        \draw (0,0) node{\color{mygreen}{$\blacksquare$} \color{black} :};
      \end{tikzpicture}
    };

    \node at (c2) {
      \begin{tikzpicture}[scale=0.3]
        \coordinate (0) at (1,1);
        \coordinate (1) at (0,0);
        \coordinate (2) at (2,0);
        
        \node[above] at (0) {\small$I_0$};
        \draw (1) -- (0) -- (2);
        \draw (0) node{\small $\bullet$};
        \draw[mygreen] (1) node{\small $\bullet$};
        \draw (2) node{\small $\bullet$};
      \end{tikzpicture}
    };

    \node at (d1) {
      \begin{tikzpicture}[scale=1]
        \draw (0,0) node{\color{blue}{$\blacksquare$} \color{black} :};
      \end{tikzpicture}
    };

    \node at (d2) {
      \begin{tikzpicture}[scale=0.3]
        \coordinate (0) at (1,1);
        \coordinate (1) at (0,0);
        \coordinate (2) at (2,0);
        
        \node[above] at (0) {\small$I_2$};
        \node[above left] at (1) {\small$I_0$};
        \draw (1) -- (0) -- (2) -- (1);
        \draw (0) node{\small $\bullet$};
        \draw[blue] (1) node{\small $\bullet$};
        \draw[blue] (2) node{\small $\bullet$};
      \end{tikzpicture}
    };

    \foreach \a in {a2,b2,c2,d2}
    \draw (\a) circle (1);

    \node at (2,0) {
      \begin{tikzpicture}[scale=1]
        \coordinate (r) at (2,2);
        \coordinate (0) at (0,1);
        \coordinate (1) at (4,1);
        \coordinate (10) at (3,0);
        \coordinate (11) at (5,0);

        \node[shape=rectangle,fill=red] at (r) {};
        \node[shape=rectangle,fill=mygreen] at (0) {};
        \node[shape=rectangle,fill=myyellow] at (1) {};
        \node[shape=rectangle,fill=mygreen] at (10) {};
        \node[shape=rectangle,fill=blue] at (11) {};

        \foreach \a/\b in {r/0,r/1,1/10,1/11} 
        \draw[->,>=latex,shorten <=3.5,shorten >=3] (\a) -- (\b);

      \end{tikzpicture}
    };

    \node at (8,0) {
      \begin{tikzpicture}[scale=1]
        \coordinate (0) at (4,4);
        \coordinate (1) at (3,3);
        \coordinate (2) at (5,3);
        \coordinate (3) at (1,3);
        \coordinate (4) at (2,2);
        \coordinate (5) at (4,2);
        \coordinate (6) at (6,2);
        \coordinate (7) at (3,1);
        \coordinate (8) at (5,1);
        
        \foreach \a/\b in {0/1,0/2,1/2,1/3,1/4,1/5,2/5,2/6,5/7,5/8,5/6} \draw (\a) -- (\b);
        \foreach \a in {4,8} \draw (\a) node{\small $\bullet$};
        \draw[mygreen] (0) node{\small $\bullet$};
        \draw[mygreen] (3) node{\small $\bullet$};
        \draw[mygreen] (7) node{\small $\bullet$};
        \draw[red] (1) node{\small $\bullet$};
        \draw[myyellow] (2) node{\small $\bullet$};
        \draw[blue] (5) node{\small $\bullet$};
        \draw[blue] (6) node{\small $\bullet$};
        
        \node at (0) [name=a0,outer sep=10pt,inner sep=10pt]{};
        \node at (1) [name=a1,outer sep=10pt,inner sep=10pt]{};
        \node at (2) [name=a2,outer sep=10pt,inner sep=10pt]{};
        \node at (3) [name=a3,outer sep=10pt,inner sep=10pt]{};
        \node at (4) [name=a4,outer sep=10pt,inner sep=10pt]{};
        \node at (5) [name=a5,outer sep=10pt,inner sep=10pt]{};
        \node at (6) [name=a6,outer sep=10pt,inner sep=10pt]{};
        \node at (7) [name=a7,outer sep=10pt,inner sep=10pt]{};
        \node at (8) [name=a8,outer sep=10pt,inner sep=10pt]{};

        \draw[rounded corners=20pt,dotted] (a0.north) -- ($(a1.south west)!0.25!(a1.north west)$) -- ($(a2.south east)!0.25!(a2.north east)$) -- cycle;
        \draw[rounded corners=20pt,dotted] (a4.south) -- ($(a1.north east)!0.25!(a1.south east)$) -- ($(a3.south west)!0.75!(a3.north west)$) -- cycle;
        \draw[rounded corners=20pt,dotted] (a5.south) -- ($(a2.north east)!0.25!(a2.south east)$) -- ($(a1.south west)!0.75!(a1.north west)$) -- cycle;
        \draw[rounded corners=20pt,dotted] (a5.north) -- ($(a7.south west)!0.25!(a7.north west)$) -- ($(a8.south east)!0.25!(a8.north east)$) -- cycle;
        \draw[rounded corners=20pt,dotted] (a2.north) -- ($(a5.south west)!0.25!(a5.north west)$) -- ($(a6.south east)!0.25!(a6.north east)$) -- cycle;
        
      \end{tikzpicture}
    };
    
  \end{tikzpicture}
  \caption{On colored graphs, $\structarbre\in\TD{2}$ (on the left) and $\ext$ (on the right)}
  \label{fig:treedec}
\end{figure}
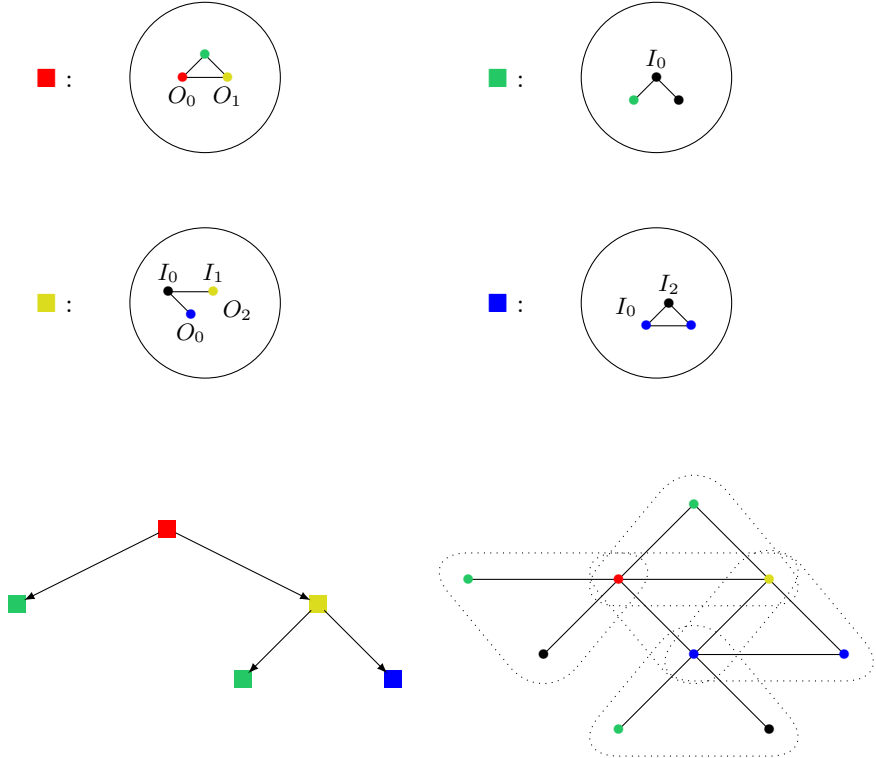

An illustration of this process is given in Figure~\ref{fig:treedec}. In this example, $k=2$ and  $\vocab=\{E,P_{\text{red}},P_{\text{yellow}},P_{\text{green}},P_{\text{blue}}\}$, i.e. $\vocab$ is the vocabulary of graphs colored with four colors, represented in the figure as colored circles. The colored squares represent the equivalence classes of $2$-bags, i.e. the unary predicates of $\vocabdec[2]$.

A \textbf{tree-decomposition} (resp. \textbf{path-decomposition}) \textbf{of width $k$} of a $\vocab$-structure $\struct$ is a $\vocabdec$-structure $\structarbre\in\TD{k}$ (resp. $\structarbre\in\PD{k}$) such that $\ext\simeq\struct$. Let $\TW{k}$ (resp. $\PW{k}$) denote the class of all $\vocab$-structures of \textbf{treewidth} (resp. \textbf{pathwidth}) \textbf{at most $k$}, that is structures that admit a tree-decomposition (resp. path-decomposition) of width at most $k$.

We say that a class $\classe$ has \textbf{treewidth} (resp. \textbf{pathwidth}) \textbf{at most $k$} if $\mathcal C\subseteq\TW{k}$ (resp. $\mathcal C\subseteq\PW{k}$), and that $\mathcal C$ is a class of \textbf{bounded treewidth} (resp. \textbf{bounded pathwidth}) if it has treewidth (resp. pathwidth) at most $k$ for some $k\in\N$.

Note that in the case of graphs, these notions are equivalent to the classical ones: in particular, a graph (seen as a $\{E\}$-structure) admits a tree-decomposition in $\TD{k}$ in the following sense iff it has treewidth $k$ in the classical sense.

\subsection{Decompositions of bounded span}

In the classical definition of a tree-decomposition of a graph, there is no restriction to the number of bags, and their relative distance in the tree, containing a given element of the graph. This carries over to the definition we gave of a tree-decomposition of a $\vocab$-structure, in that the equivalence classes for the relation $\sim$ may be composed of elements coming from $k$-bags that are arbitrarily far apart in the tree-decomposition.

For these decompositions to have an interest with respect to \FO, it is imperative that we restrict them. Indeed, \FO is not able to determine whether two nodes in two different bags correspond to the same element of the structure, as those nodes can be located arbitrarily far apart in the tree. Hence, in all generality, it is impossible to reconstruct (via an \FO-interpretation) a structure in one of its decompositions.

To avoid this issue, we consider a more restrictive notion of decomposition, based on the definition introduced independently by Ding and Oporowski~\cite{DBLP:journals/jgt/DingO95} and by Bodlaender and Engelfriet~\cite{DBLP:journals/jal/BodlaenderE97}. Let's first give the definition from \cite{DBLP:journals/jal/BodlaenderE97}.

A tree or path-decomposition (in the classical sense) of a graph is said to be domino if every element of the graph belongs to at most two bags.

If a (simple and loopless) graph admits a domino tree-decomposition of width $k$, then its degree is necessarily bounded by $2k$. The converse question of finding a domino tree-decomposition of small width for any graph of bounded degree and bounded treewidth was answered in \cite{DBLP:journals/jgt/DingO95} and \cite{DBLP:journals/jal/BodlaenderE97}. Bodlaender later improved the bound in \cite{DBLP:journals/dmtcs/Bodlaender99}: 

\begin{thm}
  \label{th:domino_dec}
  Let $k\in\N$ and $d\in\N$.

  Any graph of treewidth at most $k$ and degree at most $d$ admits a domino tree-decomposition of width at most $(9k+7)d(d+1)-1$.
\end{thm}

Let's now extend the definition of domino decompositions to the general setting of $\vocab$-structures. Our motivation is to be able to interpret in \FO a structure in its decompositions. For that, we may weaken the condition that an element can appear only in two (or a bounded number of) bags. Indeed, as long as all the bags containing a given element are in a bounded radius, \FO will be able to recreate its neighborhood in the structure; it is not necessary for their number to be bounded.

The \textbf{span} of a tree-decomposition $\structarbre\in\TW{k}$ is defined as the maximum over the equivalence classes $c$ of $\sim$ of the maximal $\dist{\structarbre}{t}{u}$, where $t,u\in\arbredom$ each contain an element of $c$. In other word, the span of $\structarbre$ is the maximal distance between two bags containing elements that are merged in $\ext$.

Let $\TDD{k}{\delta}$ (resp. $\PDD{k}{\delta}$) denote the class of tree-decompositions (resp. path-decompositions) of width at most $k$ and of span at most $\delta$, and $\TWD{k}{\delta}$ (resp. $\PWD{k}{\delta}$) denote the class of $\vocab$-structures admitting a decomposition in $\TDD{k}{\delta}$ (resp. $\PDD{k}{\delta}$).

\begin{note}
  In the case of graphs, $\TDD{k}{1}$ corresponds to all the domino tree-decompositions of width $k$.
  
  However, as soon as $\delta\geq 2$, $\TWD{k}{\delta}$ contains graphs of arbitrarily large degree. For instance, $\TWD{1}{2}$ contains all the star graphs.
\end{note}

It is straightforward to construct, for any $k,\delta\in\N$, an \FO-interpretation $\inter[k][\delta]$ such that for every $\structarbre\in\TDD{k}{\delta}$, $\inter[k][\delta](\structarbre)\simeq\ext$.

Now that we have restricted the class of decompositions, the natural question is whether it is possible to interpret in \FO a tree-decomposition of bounded span in structures of $\TWD{k}{\delta}$. In other words, does there exists \FO-interpretations $\mathcal J_k^{\delta}$ such that, for every $\mathcal A\in\TWD{k}{\delta}$, $\ext[\mathcal J_k^{\delta}(\struct)]\simeq\mathcal\struct$?

More generally, is to possible, given two \FO-similar structures, to find respective tree-decompositions of bounded span that are \FO-similar with one another? This is Question~\ref{qu:tw}. We prove in Section~\ref{sec:tw} that such decompositions do not always exist, even when we allow the width and span to be non-optimal.

As an intermediate step, which proves to be an interesting result in its own right, we show in Section~\ref{sec:pw} that Question~\ref{qu:pw}, where we look at the possibility to find \FO-continuous path-decompositions of bounded span, have a negative answer as well.

\section{Path-decompositions of bounded span are not \FO-continuous}
\label{sec:pw}

First, we investigate the case of bounded pathwidth. We give in this section a negative answer to Question~\ref{qu:pw}.

\bigskip

Let's consider the vocabulary of colored graphs $\vocab:=\{E,P_0,P_1\}$ where $E$ is a binary relation symbol, and $P_0,P_1$ are unary.

Let $k,\delta\in\N$. We exhibit two families $\counterfam$ and $\counterfambis$ of $\vocab$-structures of $\PWD{2}{2}$ such that
\begin{itemize}
\item $\forall\simindexbis\in\N,\quad\structcounter\foeq{\simindexbis}\structcounterbis$
\item $\exists\simindex\in\N,\forall\simindexbis\in\N,$ for all decompositions $\deccounter,\deccounterbis\in\PDD{k}{\delta}$,\[\deccounter\nfoeq{\simindex}\deccounterbis\,.\]
\end{itemize}

In other words, we show that no matter how large, both in term of width and span, we allow our decompositions to be, there are arbitrarily \FO-similar structures of $\PWD{2}{2}$ that do not have \FO-similar path-decompositions of this width and span. 

\bigskip

Let $k,\delta\in\N$. We set $\simindex:=\delta(n+1)$, where the value of $n$ will be apparent later on, and depends only on $k$.

Let's fix $\simindexbis\in\N$. 

\begin{figure}[!ht]
  \centering
  \begin{tikzpicture}[scale=1]

    \coordinate (0) at (0,0);
    \coordinate (1) at (5,0);
    \draw (0) node{$\bullet$} node[left]{$s$};
    \draw (1) node{$\bullet$} node[right]{$t$};
    
    \draw [dashed,thick,bend left=70,
      postaction={decorate,decoration={raise=.8ex,text along path,text align=center,text={{$pn-1$}{}}}},
      postaction={decorate,decoration={raise=-.5ex,text along path,text align/right indent=1cm,text align=right,text={{$\bullet$}{}}}},
      postaction={decorate,decoration={raise=-2.2ex,text along path,text align/right indent=.9cm,text align=right,text={{$a_1$}{}}}},
      postaction={decorate,decoration={raise=.8ex,text along path,text align/right indent=.5cm,text align=right,text={{$2^\simindexbis$}{}}}},
      postaction={decorate,decoration={raise=.8ex,text along path,text align/left indent=.5cm,text={{$2^\simindexbis$}{}}}},
      postaction={decorate,decoration={raise=-2.2ex,text along path,text align/left indent=.9cm,text={{$a_0$}{}}}},
      postaction={decorate,decoration={raise=-.5ex,text along path,text align/left indent=1cm,text={{$\bullet$}{}}}}
    ] (0) to (1);

        \draw [dashed,thick,bend right=70,
      postaction={decorate,decoration={raise=-2ex,text along path,text align=center,text={{$pn-1$}{}}}},
      postaction={decorate,decoration={raise=-.7ex,text along path,text align/right indent=1cm,text align=right,text={{$\bullet$}{}}}},
      postaction={decorate,decoration={raise=1.2ex,text along path,text align/right indent=.9cm,text align=right,text={{$a_3$}{}}}},
      postaction={decorate,decoration={raise=-2.2ex,text along path,text align/right indent=.5cm,text align=right,text={{$2^\simindexbis$}{}}}},
      postaction={decorate,decoration={raise=-2.2ex,text along path,text align/left indent=.5cm,text={{$2^\simindexbis$}{}}}},
      postaction={decorate,decoration={raise=1.2ex,text along path,text align/left indent=.9cm,text={{$a_2$}{}}}},
      postaction={decorate,decoration={raise=-.7ex,text along path,text align/left indent=1cm,text={{$\bullet$}{}}}}
    ] (0) to (1);

  \end{tikzpicture}
  \caption{The gadget \gadg{p}{n}, where the lengths of the different unoriented paths depend on $\simindexbis,p,n\in\N$.}
  \label{fig:gadget}
\end{figure}
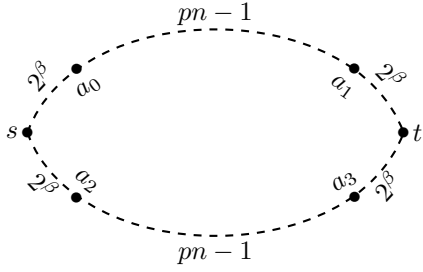

The structures $\structcounter$ and $\structcounterbis$ will be based on a series of gadgets. The first gadget, \gadg{p}{n}, is defined in Figure~\ref{fig:gadget}. The value of $p$ will be specified later on.

For any integers $n_1,n_2\leq n$, we define \bicol{p}{n}{n_1}{n_2} as \gadg{p}{n}, where the path (of length $pn-1$, hence having $pn$ nodes) from $a_0$ to $a_1$ is colored with $P_0,P_1$ as
\[(0^{n-n_1}1^{n_1})^p\]
and the path from $a_2$ to $a_3$ is colored with $P_0,P_1$ as
\[(0^{n-n_2}1^{n_2})^p\,.\]

$\vocab$ being fixed before $k$ and $\delta$ are known, these colorings are a way to encode a number of colors which can depend on $k$ and $\delta$. The integer $\simindex$, which can depend on $k$ and $\delta$, will help us decode these colorings.

\bigskip

The following lemma states that, both in path and tree-decompositions of bounded span, two bags containing nodes that are close to one another in the structure cannot be too far apart in the decomposition.

\begin{lemma}
  \label{lem:dist_dec}
  If $x$ and $y$ are two elements of a structure $\structgraphe$, then in any (path or tree) decomposition  of $\structgraphe$ of span at most $\delta$, a bag containing $x$ and a bag containing $y$ must be at distance at most $\delta\cdot(\dist{\structgraphe}{x}{y}+1)$ from each other.

\end{lemma}

  The proof of this fact is a straightforward induction on $\dist{\structgraphe}{x}{y}$.

  \bigskip

Consider a path-decomposition $\structarbre\in\PDD{k}{\delta}$ of \bicol{p}{n}{n_1}{n_2}.

Suppose that any two bags containing respectively $s$ and $t$ are at distance at least \[2\delta(2^\simindexbis+1)+1\] from one another in $\structarbre$.

Then there must exist a bag containing some node of the path $[a_0,a_1]$ as well as some node of the path $[a_2,a_3]$. Indeed, according to Lemma~\ref{lem:dist_dec}, any bag containing $a_0$ must be at distance at most $\delta(2^\simindexbis+1)$ from any bag containing $s$ and similarly for $a_2$ and $s$, $a_1$ and $t$ and $a_3$ and $t$. By assumption, there must exists at least one bag in $\structarbre$ that separates all the bags containing $a_0$ or $a_2$ from all the bags containing $a_1$ or $a_3$. Such a bag must contain both a node of $[a_0,a_1]$ and a node of $[a_2,a_3]$.

In that case, $\structarbre$ satisfies the property \suppcol{n}{n_1}{n_2}: "there exists a bag containing both a node that is part of a path $0^{n-n_1}1^{n_1}$, and a node that is part of a path $0^{n-n_2}1^{n_2}$".

Note that any path-decomposition $\structarbrebis\in\PDD{k}{\delta}$ such that $\structarbrebis\foeq{\simindex}\structarbre$ must also satisfy \suppcol{n}{n_1}{n_2}; recall that $\alpha=\delta\cdot(n+1)$.

\bigskip

We now define \bicolit{p}{n}{n_1}{n_2}{m} as the concatenation of $m$ copies of \bicol{p}{n}{n_1}{n_2}, as illustrated in Figure~\ref{fig:bicolit}.

\begin{figure}[!ht]
  \centering
  \begin{tikzpicture}[scale=2.6]

    \foreach \n in {0,...,4}{
      \coordinate (\n) at (\n,0);
      \coordinate (mid\n) at ({{\n+.5}},0);
      \draw (\n) node{$\bullet$};
    }
    
    \foreach \n in {0,1,3}{
      \draw (mid\n) node{$\bicol{p}{n}{n_1}{n_2}$};
    }

    \coordinate (dots) at (2.4,0);
    \draw (dots) node{$\cdots$};

    \draw (0) node[left]{$s_0$};
    \draw (1) node[above=1ex]{$s_1$};
    \draw (2) node[right]{$s_2$};
    \draw (3) node[left]{$s_{m-1}$};
    \draw (4) node[right]{$s_m$};

    \foreach \from/\to in {0/1,1/0,1/2,2/1,3/4,4/3}{
      \draw [-,thick,bend left=70] (\from) to (\to);
    }
      
  \end{tikzpicture}
  \caption{The gadget \bicolit{p}{n}{n_1}{n_2}{m}.}
  \label{fig:bicolit}
\end{figure}

\bigskip

Now, consider a path-decomposition $\structarbre\in\PDD{k}{\delta}$ of \bicolit{p}{n}{n_1}{n_2}{m}.

Suppose that for every $0\leq i<m$, there exist bags containing respectively $s_i$ and $s_{i+1}$ that are at distance at most \[2\delta(2^\simindexbis+1)\] from one another. Then the length of $\structarbre$ cannot be too large. Indeed, any element of \bicolit{p}{n}{n_1}{n_2}{m} is at distance at most \[2^\simindexbis+\frac{pn-1}{2}\] from the nearest $s_i$.
This means, according to Lemma~\ref{lem:dist_dec}, that the length of $\structarbre$ can be bounded by \[\delta(2^\simindexbis+\frac{pn-1}{2})+m\cdot(2\delta(2^\simindexbis+1)+\delta) +\delta(2^\simindexbis+\frac{pn-1}{2})\]
(joining a bag containing the nearest $s_i$, then jumping at most $m$ times to bags containing other $s_j$ until we reach the closest to the destination point, then reaching the destination).

This expression can be coarsely bounded by
\[\delta (2^{\simindexbis+4}m+ pn)\,.\]

This means that the size of \bicolit{p}{n}{n_1}{n_2}{m} is at most
\begin{equation}
  \label{eq:size_max}
  (k+1)[\delta (2^{\simindexbis+4}m+ pn)+1]\,.
\end{equation}
However, \gadg{p}{n} has size \[2^{\simindexbis+2}+2pn-2\,,\] hence \bicolit{p}{n}{n_1}{n_2}{m} has size \[  m(2^{\simindexbis+2}+2pn-2)-(m-1)\,,\] which is greater than
\begin{equation}
  \label{eq:size}
  m(2^{\simindexbis+2}+2pn-3)\,.
\end{equation}

The integers $p$ and $m$ can now be chosen such that the expression (\ref{eq:size}) is bigger than (\ref{eq:size_max}), which is absurd.

Indeed, showing that (\ref{eq:size}) is bigger that (\ref{eq:size_max}) amounts to finding $p$ and $m$ such that
\[m[2^{\simindexbis+2}+2pn-3-(k+1)\delta 2^{\simindexbis+4}]> (k+1)(\delta pn+1)\,.\]
This can be done by first choosing $p$ so that \[2^{\simindexbis+2}+2pn-3-(k+1)\delta 2^{\simindexbis+4}>0\,,\]
and choosing $m$ in consequence.

Hence, there must exist $0\leq i<m$ such that no pair of bags containing respectively $s_i$ and $s_{i+1}$ at distance at most $\delta\cdot 2^{\simindexbis+1}$ from each other. As we've seen, this means that $\structarbre$ satisfies \suppcol{n}{n_1}{n_2}.

\bigskip

We now have constructed a gadget \bicolit{p}{n}{n_1}{n_2}{m} which is such that any path-decomposition $\structarbre\in\PDD{k}{\delta}$ of \bicolit{p}{n}{n_1}{n_2}{m} satisfies \suppcol{n}{n_1}{n_2}.

Since \suppcol{n}{n_1}{n_2} is a property that is preserved by $\foeq{\simindex}$, we will use it as a lever to prove that there is no path-decompositions in $\PDD{k}{\delta}$ of $\structcounter$ and $\structcounterbis$ that are \FO-similar at depth $\simindex$.

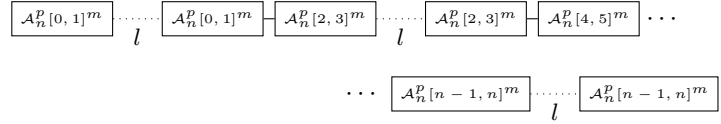
\begin{figure}[!ht]
  \centering
  \begin{tikzpicture}[scale=1]

    \node[shape=rectangle,draw=black] (0) at (0,0) {\tiny $\bicolit{p}{n}{0}{1}{m}$};
    \node[shape=rectangle,draw=black] (1) at (2,0) {\tiny $\bicolit{p}{n}{0}{1}{m}$};
    \node[shape=rectangle,draw=black] (2) at (3.5,0) {\tiny $\bicolit{p}{n}{2}{3}{m}$};
    \node[shape=rectangle,draw=black] (3) at (5.5,0) {\tiny $\bicolit{p}{n}{2}{3}{m}$};
    \node[shape=rectangle,draw=black] (4) at (7,0) {\tiny $\bicolit{p}{n}{4}{5}{m}$};
    \node (5) at (8,0) {$\cdots$};
    \node (6) at (4,-1) {$\cdots$};
    \node[shape=rectangle,draw=black] (7) at (5.3,-1) {\tiny $\bicolit{p}{n}{n-1}{n}{m}$};
    \node[shape=rectangle,draw=black] (8) at (7.8,-1) {\tiny $\bicolit{p}{n}{n-1}{n}{m}$};
    
    \path[dotted] (0) edge node[below] {\small $l$} (1);
    \path[-] (1) edge (2);
    \path[dotted] (2) edge node[below] {\small $l$} (3);
    \path[-] (3) edge (4);
    \path[dotted] (7) edge node[below] {\small $l$} (8);
     
  \end{tikzpicture}
  \caption{The structure $\structcounter$, where $p,m$ are chosen as above, and $n$ is an odd integer to be fixed. The dotted edges are unoriented paths of length $l$, which will also be fixed later on.}
  \label{fig:g_beta}
\end{figure}

For that, we define $\structcounter$ as in Figure~\ref{fig:g_beta}. The value of $n$ and $l$ will be fixed in the remainder of the proof. Recall that while $l$ will depend on $\simindexbis$, $n$ must not, since $\simindex$ itself depends on $n$.

Using the same values for $n$ and $l$, $\structcounterbis$ is defined as in Figure~\ref{fig:h_beta}.
\begin{figure}[!ht]
  \centering
  \begin{tikzpicture}[scale=1]

    \node[shape=rectangle,draw=black] (0) at (0,0) {\tiny $\bicolit{p}{n}{0}{0}{m}$};
    \node[shape=rectangle,draw=black] (1) at (2,0) {\tiny $\bicolit{p}{n}{1}{1}{m}$};
    \node[shape=rectangle,draw=black] (2) at (3.5,0) {\tiny $\bicolit{p}{n}{2}{2}{m}$};
    \node[shape=rectangle,draw=black] (3) at (5.5,0) {\tiny $\bicolit{p}{n}{3}{3}{m}$};
    \node[shape=rectangle,draw=black] (4) at (7,0) {\tiny $\bicolit{p}{n}{4}{4}{m}$};
    \node (5) at (8,0) {$\cdots$};
    \node (6) at (3.9,-1) {$\cdots$};
    \node[shape=rectangle,draw=black] (7) at (5.3,-1) {\tiny $\bicolit{p}{n}{n-1}{n-1}{m}$};
    \node[shape=rectangle,draw=black] (8) at (7.8,-1) {\tiny $\bicolit{p}{n}{n}{n}{m}$};
    
    \path[dotted] (0) edge node[below] {\small $l$} (1);
    \path[-] (1) edge (2);
    \path[dotted] (2) edge node[below] {\small $l$} (3);
    \path[-] (3) edge (4);
    \path[dotted] (7) edge node[below] {\small $l$} (8);
     
  \end{tikzpicture}
  \caption{The structure $\structcounterbis$.}
  \label{fig:h_beta}
\end{figure}

One can easily see that $\structcounter\foeq{\simindexbis}\structcounterbis$. This result from the observation that, for every $0\leq i\leq\frac{n-1}{2}$, two copies of $\bicolit{p}{n}{2i}{2i+1}{m}$ are \FO-similar at depth $\simindexbis$ to the union of $\bicolit{p}{n}{2i}{2i}{m}$ and $\bicolit{p}{n}{2i+1}{2i+1}{m}$. Indeed, the paths of length $2^\simindexbis$ in all the $\bicol{p}{n}{n_1}{n_2}$ prevent the Spoiler in the $\simindexbis$-round \EF game from spotting which integers $n_1,n_2$ appear in the same $\bicol{p}{n}{n_1}{n_2}$.

Furthermore, $\structcounter$ and $\structcounterbis$ belong to $\PWD{2}{2}$. Indeed, they each admit a path-decomposition of width $2$ that goes from their left to their right, which moves in each $\bicol{p}{n}{n_1}{n_2}$ one step in the top path, then one step in the bottom path.

\bigskip

It remains to prove that $\structcounter$ and $\structcounterbis$ do not admit decompositions in $\PDD{k}{\delta}$ that are \FO-similar at depth $\simindex$.

Suppose that $\deccounter,\deccounterbis\in\PDD{k}{\delta}$ are respective decompositions of $\structcounter$ and $\structcounterbis$ such that
\begin{equation}
  \label{eq:alpha_eq}
  \deccounter\foeq{\simindex}\deccounterbis\,.
\end{equation}

As we've seen above, $\deccounter$ must satisfy all the properties \[\suppcol{n}{0}{1},\suppcol{n}{2}{3},\cdots,\suppcol{n}{n-1}{n}\,.\]

By (\ref{eq:alpha_eq}), $\deccounterbis$ must satisfy them too. By construction of $\deccounterbis$, this means that for every $0\leq i\leq \frac{n-1}{2}$, there exists a bag of $\deccounterbis$ containing both a node of $\bicolit{p}{n}{2i}{2i}{m}$ and a node of $\bicolit{p}{n}{2i+1}{2i+1}{m}$. This prevents $\deccounterbis$ from being to long.

More precisely, since any $\bicolit{p}{n}{n_1}{n_2}{m}$ has diameter bounded by
\[m(pn-1+2^{\simindexbis+1})\,.\]
Lemma~\ref{lem:dist_dec} entails that any of its decompositions in $\PDD{k}{\delta}$ has length at most
\[\delta[m(pn-1+2^{\simindexbis+1})+1]\,.\]

With the requirement that for every $i$, $\bicolit{p}{n}{2i}{2i}{m}$ and $\bicolit{p}{n}{2i+1}{2i+1}{m}$ overlap in $\deccounterbis$, the length of $\deccounterbis$ is at most
\[\frac{l}{2}+n\delta[m(pn-1+2^{\simindexbis+1})+2]+\frac{l}{2}\,,\]
which can be bounded by
\[n\delta m(pn+2^{\simindexbis+1})+l-1\,.\]
This implies that the size of $\structcounterbis$ cannot exceed
\begin{equation}
  \label{eq:h_size_max}
  (k+1)[n\delta m(pn+2^{\simindexbis+1})+l]\,.
\end{equation}

However, recall from (\ref{eq:size}) that any \bicolit{p}{n}{n_1}{n_2}{m} has size at least
\[m(2^{\simindexbis+2}+2pn-3)\,,\]
hence the size of $\structcounterbis$ is at least
\begin{equation}
  \label{eq:h_size}
  (n+1)m(2^{\simindexbis+2}+2pn-3)+n(l-1)\,.
\end{equation}

The right choice of $n$ and $l$ make (\ref{eq:h_size}) bigger than (\ref{eq:h_size_max}), which is absurd.

Indeed, (\ref{eq:h_size}) is bigger that (\ref{eq:h_size_max}) iff
\[l[n-(k+1)]>(k+1)[n\delta m(pn+2^{\simindexbis+1})+l]-(n+1)m(2^{\simindexbis+2}+2pn-3)+n\,.\]

Choosing $n:=k+2$ allows us to set $l$ so that this inequality holds.

It follows that $\structcounter$ and $\structcounterbis$ do not have decompositions in $\PDD{k}{\delta}$ that are \FO-similar at depth $\simindex$, thus concluding the proof that the answer to Question~\ref{qu:pw} is negative.

\section{Tree-decompositions of bounded span are not \FO-continuous}
\label{sec:tw}

We have seen in Section~\ref{sec:pw} that path-decompositions of bounded span are not \FO-continuous, even when we allow one to increase the width and the span of the decompositions.

Let's now show an equivalent result for tree-decompositions by giving a negative answer to Question~\ref{qu:tw}. For that, we proceed in a somewhat similar way (we also try to constrain the size of the decompositions), although the construction is significantly more involved.

\bigskip

Once again, let's consider the vocabulary of colored graphs $\vocab:=\{E,P_0,P_1\}$ where $E$ is binary and $P_0,P_1$ are unary.

Let $k,\delta\in\N$. We exhibit two families $\counterfam$ and $\counterfambis$ of $\vocab$-structures of $\TW{2}$ and of degree $5$ (hence, by Theorem~\ref{th:domino_dec}, belonging to $\TWD{749}{1}$) such that

\begin{itemize}
\item $\forall\simindexbis\in\N,\quad\structcounter\foeq{\simindexbis}\structcounterbis$
\item $\exists\simindex\in\N,\forall\simindexbis\in\N,$ for all decompositions $\deccounter,\deccounterbis\in\TDD{k}{\delta}$,\[\deccounter\foneq{\simindex}\deccounterbis\,.\]
\end{itemize}

\begin{figure}[!ht]
  \centering
  \begin{tikzpicture}[scale=1]

    \coordinate (c) at (5.5,0);
    \draw (c) node{$\cdots$};

    \coordinate (att) at (-1,4.25);
    \coordinate (at) at (-1,1.5);
    \coordinate (abb) at (-1,-4.25);
    \coordinate (ab) at (-1,-1.5);

    \draw[<->] (att) -- node[midway,left]{$p$} (at);
    \draw[<->] (ab) -- node[midway,left]{$l$} (at);
    \draw[<->] (abb) -- node[midway,left]{$p$} (ab);

    \foreach \n in {0,...,7}{
      \coordinate (ttt\n) at ({{\n+.5}},2);
      \coordinate (tt\n) at (\n,1.5);
      \coordinate (t\n) at (\n,.5);
      \coordinate (\n) at (\n,0.15);
      \coordinate (b\n) at (\n,-.5);
      \coordinate (bb\n) at (\n,-1.5);
      \coordinate (bbb\n) at ({{\n+.5}},-2);

      \foreach \from/\to in {tt\n/t\n,bb\n/b\n}{
        \draw[dashed] (\from) to (\to);
      }
      \foreach \a in {tt\n,bb\n}{
        \draw (\a) node{$\bullet$};
      }
      
      \draw (\n) node{$\vdots$};
    }

    \foreach \n/\a in {0/0,1/1,2/2,3/3,4/4,5/5,6/2^p-2}{
      \draw (tt\n) node[left]{\small $a_{\a}$};
      \draw (bb\n) node[left]{\small $b_{\a}$};
    }

    \foreach \n/\a in {7/2^p-1}{
      \draw (tt\n) node[right]{\small $a_{\a}$};
      \draw (bb\n) node[right]{\small $b_{\a}$};
    }

    \foreach \n in {0,...,3}{
      \coordinate (ttt\n) at ({{2*\n+.5}},2);
      \coordinate (bbb\n) at ({{2*\n+.5}},-2);

      \coordinate (tttt\n) at ({{2*\n+.5}},2.7);
      \coordinate (bbbb\n) at ({{2*\n+.5}},-2.5);

      \coordinate (t5\n) at ({{\n+2}},3.25);
      \coordinate (b5\n) at ({{\n+2}},-3.25);
    }
    
    \draw (tttt0) node[right=10pt]{$\iddots$};
    \draw (tttt1) node{$\vdots$};
    \draw (tttt2) node{$\vdots$};
    \draw (tttt3) node[left=10pt]{$\ddots$};

    \draw (bbbb0) node[right=5pt]{$\ddots$};
    \draw (bbbb1) node{$\vdots$};
    \draw (bbbb2) node{$\vdots$};
    \draw (bbbb3) node[left=5pt]{$\iddots$};

    \coordinate (t60) at (2.5,3.75);
    \coordinate (t61) at (4.5,3.75);

    \coordinate (t7) at (3.5,4.25);

    \coordinate (b60) at (2.5,-3.75);
    \coordinate (b61) at (4.5,-3.75);

    \coordinate (b7) at (3.5,-4.25);

    \foreach \from/\to in {t50/t60,t51/t60,t52/t61,t53/t61,t60/t7,t61/t7,b50/b60,b51/b60,b52/b61,b53/b61,b60/b7,b61/b7}{
      \draw[-] (\from) to (\to);
    }

    \foreach \a in {ttt0,ttt1,ttt2,ttt3,bbb0,bbb1,bbb2,bbb3,t50,t51,t52,t53,t60,t61,t7,b50,b51,b52,b53,b60,b61,b7}{
      \draw (\a) node{$\bullet$};
    }

    \foreach \from/\to in {tt0/ttt0,tt1/ttt0,tt2/ttt1,tt3/ttt1,tt4/ttt2,tt5/ttt2,tt6/ttt3,tt7/ttt3,bb0/bbb0,bb1/bbb0,bb2/bbb1,bb3/bbb1,bb4/bbb2,bb5/bbb2,bb6/bbb3,bb7/bbb3}{
      \draw[-] (\from) to (\to);
    }

    \draw (t7) node[above]{\small $a$};
    \draw (b7) node[below]{\small $b$};
    
  \end{tikzpicture}
  \caption{The gadget \loz{p}{l}, where $l$ is the length of the dotted paths. All the edges are undirected. The nodes $a$ and $b$ are the \textbf{sources} of \loz{p}{l}.}
  \label{fig:loz}
\end{figure}
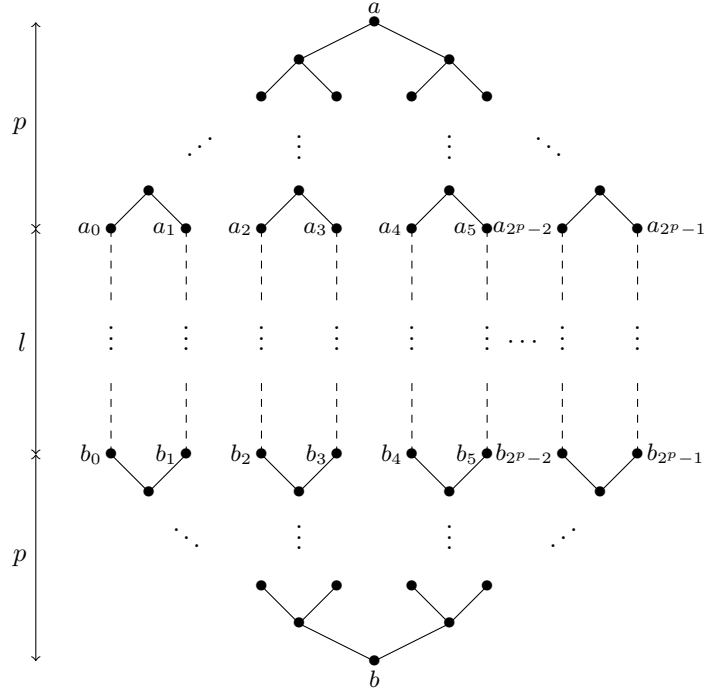

We introduce in Figure~\ref{fig:loz} the gadget \loz{p}{l}, which is composed of two complete binary trees of height $p$, whose leaves are pairwise linked by a path of length $l$. 

The interest of \loz{p}{l} resides in the fact that, provided that $p$ is large enough with respect to the width of a tree-decomposition of \loz{p}{l}, the sources (that is, the elements $a$ and $b$) cannot be far apart in this decomposition, while they can be made arbitrarily distant, by choice of $l$, in the original graph.

More precisely, let $\structarbre$ be a tree-decomposition of \loz{p}{l} in $\TDD{k}{\delta}$, where \[p=\lceil\log(k+2)\rceil\,.\] We claim that in $\structarbre$, any bag containing $a$ and any bag containing $b$ are at distance at most $d:=2\delta(p+2)+2\delta$ from one another.

Suppose otherwise, and consider the bags $t_a$ and $t_b\in T$ which minimize $\dist{\mathcal T}{t_a}{t_b}$ among bags containing respectively $a$ and $b$. Since $\mathcal T$ has span $\delta$, we must have that \[\dist{\mathcal T}{t_a}{t_b}\geq 2\delta(p+1)+1\,.\]
Let $t\in\arbredom$ be a bag on the path between $t_a$ and $t_b$ at distance at least $\delta(p+1)+1$ from each other. By virtue of $\structarbre$ being a tree and in view of Lemma~\ref{lem:dist_dec}, $t$ disconnects all the bags containing the $a_i$ from all the bags containing the $b_i$.

This means that all the $2^p$ disjoint paths from $a_i$ to $b_i$ must intersect $t$, which is absurd since $2^p\geq k+2$.

This proves that any bag containing $a$ and any bag containing $b$ are at distance at most $2\delta(p+2)$ from one another in $\structarbre$.

\bigskip

Once $k$ and $\delta$ are given, we set $p:=\lceil\log(k+2)\rceil$.

Our construction will depend on an integer $n$ whose value depends only on $k$ and $\delta$, and will be set later on. The integer $\simindex$ will also be chosen later on.

Let $\simindexbis\in\N$, and set $l:=2^\simindexbis$.

\bigskip

Let's now construct the structures $\structcounter$ and $\structcounterbis$. Both of them will amount to a concatenation of many instances of \loz{p}{l} (by concatenation of two \loz{p}{l}, we mean the disjoint union of those structures, where we merge one of their sources). On top of that, all the sources will have a label. As in Section~\ref{sec:pw}, labels are encoded with the unary relations $P_0, P_1$ on a path of length at most $n$.

\bigskip

Let's deal with $\structcounter$ first. We start by considering some nodes $(s_w)_{w\in\{0,1\}^{\leq n}}$, i.e. one element for each one of the $2^{n+1}-1$ sequences of bits of length at most $n$. All those $s_w$ belong to $\structcounter$.

We now add the aforementioned labels: for every $w\in\{0,1\}^{\leq n}$ we attach to $s_w$ a path of length $|w|$, and color it with $P_0$ and $P_1$ in order to code $w$.

On top of that, for every $w\in\{0,1\}^{\leq n-1}$, we link $s_w$ to $s_{w0}$ with an copy of \loz{p}{l}; in other words, $s_w$ and $s_{w0}$ are the sources of this gadget. All those copies are disjoint. Note that up until now, each $s_{w1}$ with $|w|=n-1$ is alone with its label in its connected component of $\structcounter$. This stage of the construction is depicted in Figure~\ref{fig:g_beta_start}.

\begin{figure}[ht!]
  \centering
  \begin{tikzpicture}[scale=2.6]

    \foreach \n in {0,...,4}{
      \coordinate (\n) at (\n,0);
      \coordinate (mid\n) at ({{\n+.5}},0);
      \draw (\n) node{$\bullet$};
    }
    
    \foreach \n in {0,1,3}{
      \draw (mid\n) node{\loz{p}{l}};
    }
    
    \coordinate (dots) at (2.5,0);
    \draw (dots) node{$\cdots$};
    
    \draw (0) node[above=1ex]{$s_\epsilon$};

    \draw (1) node[above=1ex]{$s_0$};
    \coordinate (11) at (1,-.2);
    \draw (1) -- (11);
    \draw (11) node[fill=white]{\tiny $0$};

    \draw (2) node[above=1ex]{$s_{00}$};
    \coordinate (21) at (2,-.2);
    \coordinate (22) at (2,-.4);
    \draw (2) -- (21);
    \draw (21) -- (22);
    \draw (21) node[fill=white]{\tiny $0$};
    \draw (22) node[fill=white]{\tiny $0$};

    \draw (3) node[above=1ex]{$s_{0^{n-1}}$};
    \coordinate (31) at (3,-.2);
    \coordinate (32) at (3,-.4);
    \coordinate (33) at (3,-.6);
    \draw (3) -- (31);
    \draw (31) -- (32);
    \draw[dotted] (32) -- (33);
    \draw (31) node[fill=white]{\tiny $0$};
    \draw (32) node[fill=white]{\tiny $0$};
    \draw (33) node[fill=white]{\tiny $0$};
    \draw [decorate,decoration={brace,amplitude=2pt,mirror,raise=4pt},yshift=0pt] (31) -- (33) node [midway,xshift=-0.6cm] {\tiny $n-1$};

    \draw (4) node[above=1ex]{$s_{0^n}$};
    \coordinate (41) at (4,-.2);
    \coordinate (42) at (4,-.4);
    \coordinate (43) at (4,-.6);
    \draw (4) -- (41);
    \draw (41) -- (42);
    \draw[dotted] (42) -- (43);
    \draw (41) node[fill=white]{\tiny $0$};
    \draw (42) node[fill=white]{\tiny $0$};
    \draw (43) node[fill=white]{\tiny $0$};
    \draw [decorate,decoration={brace,amplitude=2pt,mirror,raise=4pt},yshift=0pt] (41) -- (43) node [midway,xshift=-0.5cm] {\tiny $n$};

    \foreach \from/\to in {0/1,1/0,1/2,2/1,3/4,4/3}{
      \draw [-,thick,bend left=30] (\from) to (\to);
    }

    \coordinate (b1) at (.5,-1);
    \draw (b1) node{$\bullet$};
    \coordinate (b2) at (1.5,-1);
    \draw (b2) node{$\bullet$};
    \coordinate (midb1) at (1,-1);
    \foreach \n in {3,4}{
      \coordinate (b\n) at (\n,-1);
      \coordinate (midb\n) at ({{\n+.5}},-1);
      \draw (b\n) node{$\bullet$};
    }
    
    \foreach \n in {1,3}{
      \draw (midb\n) node{\loz{p}{l}};
    }

    \coordinate (bdots) at (2.25,-1);
    \draw (bdots) node{$\cdots$};
    
    \draw (b1) node[above=1ex]{$s_1$};
    \coordinate (b11) at (.5,-1.2);
    \draw (b1) -- (b11);
    \draw (b11) node[fill=white]{\tiny $1$};

    \draw (b2) node[above=1ex]{$s_{10}$};
    \coordinate (b21) at (1.5,-1.2);
    \coordinate (b22) at (1.5,-1.4);
    \draw (b2) -- (b21);
    \draw (b21) -- (b22);
    \draw (b21) node[fill=white]{\tiny $1$};
    \draw (b22) node[fill=white]{\tiny $0$};

    \draw (b3) node[above=1ex]{$s_{10^{n-2}}$};
    \coordinate (b31) at (3,-1.2);
    \coordinate (b32) at (3,-1.4);
    \coordinate (b33) at (3,-1.6);
    \draw (b3) -- (b31);
    \draw (b31) -- (b32);
    \draw[dotted] (b32) -- (b33);
    \draw (b31) node[fill=white]{\tiny $1$};
    \draw (b32) node[fill=white]{\tiny $0$};
    \draw (b33) node[fill=white]{\tiny $0$};
    \draw [decorate,decoration={brace,amplitude=2pt,mirror,raise=4pt},yshift=0pt] (b31) -- (b33) node [midway,xshift=-0.6cm] {\tiny $n-1$};

    \draw (b4) node[above=1ex]{$s_{10^{n-1}}$};
    \coordinate (b41) at (4,-1.2);
    \coordinate (b42) at (4,-1.4);
    \coordinate (b43) at (4,-1.6);
    \draw (b4) -- (b41);
    \draw (b41) -- (b42);
    \draw[dotted] (b42) -- (b43);
    \draw (b41) node[fill=white]{\tiny $1$};
    \draw (b42) node[fill=white]{\tiny $0$};
    \draw (b43) node[fill=white]{\tiny $0$};
    \draw [decorate,decoration={brace,amplitude=2pt,mirror,raise=4pt},yshift=0pt] (b41) -- (b43) node [midway,xshift=-0.5cm] {\tiny $n$};

    \foreach \from/\to in {b1/b2,b2/b1,b3/b4,b4/b3}{
      \draw [-,thick,bend left=30] (\from) to (\to);
    }

    \foreach \n in {1,...,4}{
      \coordinate (b'\n) at (\n,-2);
      \coordinate (midb'\n) at ({{\n+.5}},-2);
      \draw (b'\n) node{$\bullet$};
    }
    
    \foreach \n in {1,3}{
      \draw (midb'\n) node{\loz{p}{l}};
    }

    \coordinate (b'dots) at (2.5,-2);
    \draw (b'dots) node{$\cdots$};
    
    \draw (b'1) node[above=1ex]{$s_{01}$};
    \coordinate (b'11) at (1,-2.2);
    \coordinate (b'12) at (1,-2.4);
    \draw (b'1) -- (b'11);
    \draw (b'11) -- (b'12);
    \draw (b'11) node[fill=white]{\tiny $0$};
    \draw (b'12) node[fill=white]{\tiny $1$};

    \draw (b'2) node[above=1ex]{$s_{010}$};
    \coordinate (b'21) at (2,-2.2);
    \coordinate (b'22) at (2,-2.4);
    \coordinate (b'23) at (2,-2.6);
    \draw (b'2) -- (b'21);
    \draw (b'21) -- (b'22);
    \draw (b'22) -- (b'23);
    \draw (b'21) node[fill=white]{\tiny $0$};
    \draw (b'22) node[fill=white]{\tiny $1$};
    \draw (b'23) node[fill=white]{\tiny $0$};

    \draw (b'3) node[above=1ex]{$s_{010^{n-3}}$};
    \coordinate (b'31) at (3,-2.2);
    \coordinate (b'32) at (3,-2.4);
    \coordinate (b'33) at (3,-2.6);
    \coordinate (b'34) at (3,-2.8);
    \draw (b'3) -- (b'31);
    \draw (b'31) -- (b'32);
    \draw (b'32) -- (b'33);
    \draw[dotted] (b'33) -- (b'34);
    \draw (b'31) node[fill=white]{\tiny $0$};
    \draw (b'32) node[fill=white]{\tiny $1$};
    \draw (b'33) node[fill=white]{\tiny $0$};
    \draw (b'34) node[fill=white]{\tiny $0$};
    \draw [decorate,decoration={brace,amplitude=2pt,mirror,raise=4pt},yshift=0pt] (b'31) -- (b'34) node [midway,xshift=-0.6cm] {\tiny $n-1$};

    \draw (b'4) node[above=1ex]{$s_{010^{n-2}}$};
    \coordinate (b'41) at (4,-2.2);
    \coordinate (b'42) at (4,-2.4);
    \coordinate (b'43) at (4,-2.6);
    \coordinate (b'44) at (4,-2.8);
    \draw (b'4) -- (b'41);
    \draw (b'41) -- (b'42);
    \draw (b'42) -- (b'43);
    \draw[dotted] (b'43) -- (b'44);
    \draw (b'41) node[fill=white]{\tiny $0$};
    \draw (b'42) node[fill=white]{\tiny $1$};
    \draw (b'43) node[fill=white]{\tiny $0$};
    \draw (b'44) node[fill=white]{\tiny $0$};
    \draw [decorate,decoration={brace,amplitude=2pt,mirror,raise=4pt},yshift=0pt] (b'41) -- (b'44) node [midway,xshift=-0.5cm] {\tiny $n$};

    \foreach \from/\to in {b'1/b'2,b'2/b'1,b'3/b'4,b'4/b'3}{
      \draw [-,thick,bend left=30] (\from) to (\to);
    }

    \coordinate (vdots) at (3.5,-2.8);
    \draw (vdots) node{$\vdots$};

    \foreach \n in {3,4}{
      \coordinate (bb\n) at (\n,-3.5);
      \coordinate (midbb\n) at ({{\n+.5}},-3.5);
      \draw (bb\n) node{$\bullet$};
    }
    
    \draw (midbb3) node{\loz{p}{l}};

    \draw (bb3) node[above=1ex]{$s_{1^{n-1}}$};
    \coordinate (bb31) at (3,-3.7);
    \coordinate (bb32) at (3,-3.9);
    \coordinate (bb33) at (3,-4.1);
    \draw (bb3) -- (bb31);
    \draw[dotted] (bb31) -- (bb32);
    \draw (bb32) -- (bb33);
    \draw (bb31) node[fill=white]{\tiny $1$};
    \draw (bb32) node[fill=white]{\tiny $1$};
    \draw (bb33) node[fill=white]{\tiny $1$};
    \draw [decorate,decoration={brace,amplitude=2pt,mirror,raise=4pt},yshift=0pt] (bb31) -- (bb33) node [midway,xshift=-0.6cm] {\tiny $n-1$};

    \draw (bb4) node[above=1ex]{$s_{1^{n-1}0}$};
    \coordinate (bb41) at (4,-3.7);
    \coordinate (bb42) at (4,-3.9);
    \coordinate (bb43) at (4,-4.1);
    \draw (bb4) -- (bb41);
    \draw[dotted] (bb41) -- (bb42);
    \draw (bb42) -- (bb43);
    \draw (bb41) node[fill=white]{\tiny $1$};
    \draw (bb42) node[fill=white]{\tiny $1$};
    \draw (bb43) node[fill=white]{\tiny $0$};
    \draw [decorate,decoration={brace,amplitude=2pt,mirror,raise=4pt},yshift=0pt] (bb41) -- (bb43) node [midway,xshift=-0.5cm] {\tiny $n$};

    \foreach \from/\to in {bb3/bb4,bb4/bb3}{
      \draw [-,thick,bend left=30] (\from) to (\to);
    }

    \foreach \n in {0,1,2,3}{
      \coordinate (i\n) at (\n,-4.5);
      \draw (i\n) node{$\bullet$};
    }
    
    \coordinate (idots) at (2.5,-4.5);
    \draw (idots) node{$\cdots$};

    \draw (i0) node[above=1ex]{$s_{0^{n-1}1}$};
    \coordinate (i01) at (0,-4.7);
    \coordinate (i02) at (0,-4.9);
    \coordinate (i03) at (0,-5.1);
    \coordinate (i04) at (0,-5.3);
    \draw (i0) -- (i01);
    \draw[dotted] (i01) -- (i02);
    \draw (i02) -- (i03);
    \draw (i03) -- (i04);
    \draw (i01) node[fill=white]{\tiny $0$};
    \draw (i02) node[fill=white]{\tiny $0$};
    \draw (i03) node[fill=white]{\tiny $0$};
    \draw (i04) node[fill=white]{\tiny $1$};
    \draw [decorate,decoration={brace,amplitude=2pt,mirror,raise=4pt},yshift=0pt] (i01) -- (i04) node [midway,xshift=-0.5cm] {\tiny $n$};

    \draw (i1) node[above=1ex]{$s_{0^{n-2}11}$};
    \coordinate (i11) at (1,-4.7);
    \coordinate (i12) at (1,-4.9);
    \coordinate (i13) at (1,-5.1);
    \coordinate (i14) at (1,-5.3);
    \draw (i1) -- (i11);
    \draw[dotted] (i11) -- (i12);
    \draw (i12) -- (i13);
    \draw (i13) -- (i14);
    \draw (i11) node[fill=white]{\tiny $0$};
    \draw (i12) node[fill=white]{\tiny $0$};
    \draw (i13) node[fill=white]{\tiny $1$};
    \draw (i14) node[fill=white]{\tiny $1$};
    \draw [decorate,decoration={brace,amplitude=2pt,mirror,raise=4pt},yshift=0pt] (i11) -- (i14) node [midway,xshift=-0.5cm] {\tiny $n$};

    \draw (i2) node[above=1ex]{$s_{0^{n-3}101}$};
    \coordinate (i21) at (2,-4.7);
    \coordinate (i22) at (2,-4.9);
    \coordinate (i23) at (2,-5.1);
    \coordinate (i24) at (2,-5.3);
    \draw (i2) -- (i21);
    \draw[dotted] (i21) -- (i22);
    \draw (i22) -- (i23);
    \draw (i23) -- (i24);
    \draw (i21) node[fill=white]{\tiny $0$};
    \draw (i22) node[fill=white]{\tiny $1$};
    \draw (i23) node[fill=white]{\tiny $0$};
    \draw (i24) node[fill=white]{\tiny $1$};
    \draw [decorate,decoration={brace,amplitude=2pt,mirror,raise=4pt},yshift=0pt] (i21) -- (i24) node [midway,xshift=-0.5cm] {\tiny $n$};

    \draw (i3) node[above=1ex]{$s_{1^n}$};
    \coordinate (i31) at (3,-4.7);
    \coordinate (i32) at (3,-4.9);
    \coordinate (i33) at (3,-5.1);
    \coordinate (i34) at (3,-5.3);
    \draw (i3) -- (i31);
    \draw[dotted] (i31) -- (i32);
    \draw (i32) -- (i33);
    \draw (i33) -- (i34);
    \draw (i31) node[fill=white]{\tiny $1$};
    \draw (i32) node[fill=white]{\tiny $1$};
    \draw (i33) node[fill=white]{\tiny $1$};
    \draw (i34) node[fill=white]{\tiny $1$};
    \draw [decorate,decoration={brace,amplitude=2pt,mirror,raise=4pt},yshift=0pt] (i31) -- (i34) node [midway,xshift=-0.5cm] {\tiny $n$};

  \end{tikzpicture}
  \caption{$\structcounter$, at the beginning of the construction.}
  \label{fig:g_beta_start}
\end{figure}

\begin{figure}[ht!]
  \centering
  \begin{tikzpicture}[scale=2.6]

    \foreach \n in {0,...,4}{
      \coordinate (\n) at (\n,0);
      \coordinate (mid\n) at ({{\n+.5}},0);
      \draw (\n) node{$\bullet$};
    }
    
    \foreach \n in {0,1,3}{
      \draw (mid\n) node{\loz{p}{l}};
    }
    
    \coordinate (dots) at (2.5,0);
    \draw (dots) node{$\cdots$};
    
    \draw (0) node[above=1ex]{$s_{w_0}$};
    \coordinate (b0) at (0,-.45);
    \draw[thick,dashed] (0) -- node[near end,left]{\tiny $w_0$} (b0);

    \draw (1) node[above=1ex]{$s_{w_1}$};
    \coordinate (b1) at (1,-.45);
    \draw[thick,dashed] (1) -- node[near end,left]{\tiny $w_1$} (b1);

    \draw (2) node[above=1ex]{$s_{w_2}$};
    \coordinate (b2) at (2,-.45);
    \draw[thick,dashed] (2) -- node[near end,left]{\tiny $w_2$} (b2);

    \draw (3) node[above=1ex]{$s_{w_{2^{n+1}-3}}$};
    \coordinate (b3) at (3,-.45);
    \draw[thick,dashed] (3) -- node[near end,left]{\tiny $w_{2^{n+1}-3}$} (b3);

    \draw (4) node[above=1ex]{$s_{w_{2^{n+1}-2}}$};
    \coordinate (b4) at (4,-.45);
    \draw[thick,dashed] (4) -- node[near end,left]{\tiny $w_{2^{n+1}-2}$} (b4);

    \foreach \from/\to in {0/1,1/0,1/2,2/1,3/4,4/3}{
      \draw [-,thick,bend left=30] (\from) to (\to);
    }

  \end{tikzpicture}
  \caption{$\structcounter$, at the end of the construction, for some ordering $w_0,\cdots,w_{2^{n+1}-2}$ of the words of $\{0,1\}^{\leq n}$ to be fixed later on. It extends the construction from Figure~\ref{fig:g_beta_start}.}
  \label{fig:g_beta_end}
\end{figure}
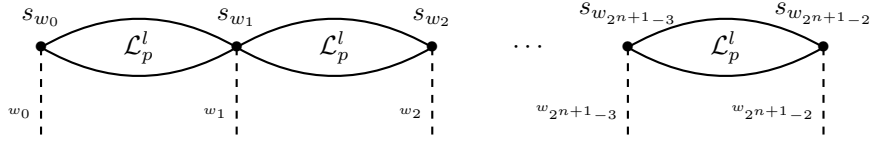

As of now, $\structcounter$ is a union of concatenations of copies of \loz{p}{l}, together with some isolated nodes. However, we will specify later how to agglomerate all those connected components so that $\structcounter$ becomes a single concatenation of copies of \loz{p}{l}. Basically, we will add copies of \loz{p}{l} between the connected components to group them into a single sequence of \loz{p}{l}, as illustrated in Figure~\ref{fig:g_beta_end}.

\bigskip

Similarly, $\structcounterbis$ is obtained by linking each $s_w$ to the corresponding $s_{w1}$ with a copy of \loz{p}{l}. In the end $\structcounterbis$ will also be a single concatenation of copies of \loz{p}{l}.


The idea behind those constructions is that, in the complete binary tree where the nodes are the words of $\{0,1\}^{\leq n}$, each edge $(w,w')$ is taken into account as a copy of \loz{p}{l} either in $\structcounter$ (if $w'=w0$) or in $\structcounterbis$ (if $w'=w1$).

The labels will help us identify $s_w$ in $\structcounter$ and $\structcounterbis$, with the help of the \FO-similarity at depth $\simindex$ of their tree-decompositions.

\bigskip

We can now start establishing some results. Suppose that $\deccounter$ and $\deccounterbis$ are respective tree-decompositions of $\structcounter$ and $\structcounterbis$ in $\TDD{k}{\delta}$ such that
\begin{equation}
  \label{eq:simdec}
  \deccounter\foeq{\simindex}\deccounterbis\,.
\end{equation}

We've seen that in $\deccounter$, for any $w\in\{0,1\}^{\leq n-1}$, any bags containing respectively $s_w$ and $s_{w0}$ are at distance at most $d:=2\delta(p+2)$ from one another. Let's choose a large enough $\simindex$ with respect to $n$, $d$ and $\delta$ so that the property \[\text{``any bags containing nodes with labels $w$ and $w0$ are at distance at most $d$''}\] is expressible as an \FO-sentence of quantifier rank $\alpha$. By (\ref{eq:simdec}), any bags of $\deccounterbis$ containing respectively $s_w$ and $s_{w0}$ must be at distance at most $d$.


Similarly, both in $\deccounter$ and $\deccounterbis$, any bags containing respectively $s_w$ and $s_{w1}$ must be at distance at most $d=2\delta(p+2)$ from one another.

In the end, for $w,w'\in\{0,1\}^{\leq n}$, in $\deccounter$ as well as in $\deccounterbis$, any bags containing respectively $s_w$ and $s_{w'}$ must be at distance at most $2nd$ from one another. This is because the complete binary tree of height $n$ as diameter $2n$.

\bigskip

The next step is to show that we can identify the parts of $\deccounter$ and $\deccounterbis$ which contain the elements $s_w$.

More precisely, we show that there exist subtrees $\mathcal S_G$ and $\mathcal S_H$ of $\deccounter$ and $\deccounterbis$ such that

\begin{itemize}
\item $\mathcal S_G\simeq \mathcal S_H$
\item the diameter of $\mathcal S_G,\mathcal S_H$ is at most $2nd$
\item $\mathcal S_G$ and $\mathcal S_H$ have degree at most $2k+3$
\item every $s_w$ belongs to at least one bag of $\mathcal S_G$, and one bag of $\mathcal S_H\,.$
\end{itemize}

For that, we start by considering the minimal subtree $\mathcal S_G$ of $\deccounter$ which contains all the bags containing any $s_w$. As we've seen, this subtree has diameter at most $2nd$.

There is however no restriction on the degree of $\mathcal S_G$. To get the desired properties, we trim $\mathcal S_G$ in the following way.

While there exists at least one, pick a bag $t$ of $\mathcal S_G$ of degree greater than $2k+3$. Let $\mathcal S_1,\cdots,\mathcal S_r$ be the connected components of $\mathcal S_G\setminus\{t\}$.

We claim that at most $2k+3$ of the $\mathcal S_i$ contain some $s_w$ which does not appear in $t$. Recall, although at this point we have only partially constructed $\structcounter$, that in the end it will be a concatenation of \loz{p}{l} (i.e. will consist of all the $s_w$, arranged in some sequence, and pairwise linked with a copy of \loz{p}{l}).

Let $w_0,\cdots,w_{2^{n+1}-2}$ be the sequence of words of $\{0,1\}^{\leq n}$ appearing in the same order as the $s_w$ in $\structcounter$, c.f. Figure~\ref{fig:g_beta_end}.

Suppose that there are at least $2k+4$ connected components $\mathcal S_i$ containing some $s_w$ which does not appear in $t$.

For each such $\mathcal S_i$, let $m(i)$ be the maximal index such that $s_{w_{m(i)}}$ belongs to $\mathcal S_i$ and not to $t$. For at most one $\mathcal S_i$ we can have $m(i)=2^{n+1}-2$. For all the others (i.e. for at least $2k+3$ of them), $s_{w_{m(i)+1}}$ does not belong to $\mathcal S_i$.

Given that $t$ is a bag of size at most $k+1$, there must exist at least $k+2$ indexes $i$ such that
\begin{itemize}
\item $s_{w_{m(i)}}$ belongs to $\mathcal S_i$ and not to $t$,
\item $s_{w_{m(i)+1}}$ belongs neither to $\mathcal S_i$ neither to $t$.
\end{itemize}

By construction, $s_{w_{m(i)}}$ and $s_{w_{m(i)+1}}$ are linked with a copy of \loz{p}{l} in $\structcounter$. Thus for each of these $k+2$ couples, there exists a path from $s_{w_j}$ to $s_{w_{j+1}}$, which must intersect $t$. All such path being disjoint, $t$ must intersect $k+2$ distinct paths, which is absurd. 

We trim out of $\mathcal S_G$ all the $\mathcal S_i$ which contain no $s_w$ which does not appear in $t$. In the new subtree, $t$ has degree at most $2k+3$, and each $s_w$ still belongs to at least one of its bags.

In the end, $\mathcal S_G$ has degree at most $2k+3$, and its diameter is a most $2nd$.

We find $\mathcal S_H$ using (\ref{eq:simdec}): by setting $\alpha$ big enough wrt. $n,k$ and $\delta$, the Spoiler in the $\simindex$-round \EF game between $\deccounter$ and $\deccounterbis$ can cover $\mathcal S_G$ (which has bounded diameter and degree) as well as enough bags of $\deccounter$ to cover the labels of each of the $s_w$. The corresponding moves of the Duplicator in $\deccounterbis$ yield $\mathcal S_H$.

In the remainder of the proof, we let $\mathcal S:=\mathcal S_G\simeq\mathcal S_H$.

\bigskip

To conclude the proof, we are now going to show that these decompositions are too compact to exist.

In $\mathcal S$, we pick a node $t$. Let $\mathcal S_1,\cdots,\mathcal S_r$ be the connected components of $\mathcal S\setminus\{t\}$, with $r\leq 2k+3$.

Let's establish some vocabulary. We say that \textbf{$s_w$ only occurs in $\mathcal S_i$}, which we denote $s_w\oocc \mathcal S_i$, if $s_w$ is contained in some bag of $\mathcal S_i$, but not in $t$. Note that this implies, by nature of tree-decompositions, that $s_w$ doesn't belong to any bag of any $\mathcal S_j,j\neq i$.

If $s_w\oocc \mathcal S_i$, we say that $s_w$ is an \textbf{$\mathcal S_i$-inode}. If $s_{w'}\oocc \mathcal S_j$ for some $j\neq i$, we say that $s_w$ is an \textbf{$\mathcal S_i$-onode}.

We say that $s_w$ and $s_{w'}$ are \textbf{adjacent in $\structcounter$} (resp. \textbf{in $\structcounterbis$}) if they are linked by a copy of \loz{p}{l} in $\structcounter$ (resp. $\structcounterbis$).

We say that they are \textbf{adjacent} if they are adjacent in $\structcounter$ or in $\structcounterbis$. For now, $s_w$ and $s_{w'}$ are adjacent iff $w'=w0$, $w'=w1$, $w=w'0$ or $w=w'1$, but as said previously, we are going to add some copies of \loz{p}{l} in both structures.

We say that $\{s_w,s_{w'}\}$ is an \textbf{$\mathcal S_i$-bridge} if
\begin{itemize}
\item $s_w\oocc \mathcal S_i$
\item $s_{w'}\oocc \mathcal S_j$ for some $j\neq i$
\item $s_w$ and $s_{w'}$ are adjacent.
\end{itemize}

Let $\mathcal S_i$ be a connected component of $\mathcal S\setminus\{t\}$. Then there are at most $2k+2$ $\mathcal S_i$-bridges.
Otherwise there would exist at least $k+2$ $\mathcal S_i$-inode adjacent in $\structcounter$ (without loss of generality) to some $\mathcal S_i$-onode. Thus $t$ would intersect with at least $k+2$ disjoint paths, which is absurd.

Let $h<n$ whose value will be apparent later on. With this remark in mind, let's now show that there cannot exist $i$ such that
\begin{equation}
  \label{eq:enc_size}
  N\leq |\{s_w\oocc \mathcal S_i\}|\leq (2^{n+1}-1)-(k+1)-N\,,
\end{equation}
where \[N:=(2^h-1)+(4k+4)(2^{n-h+1}-1)+1\,.\]
This amount to saying that the $s_w$ cannot be spread evenly across the $\mathcal S_1,\cdots,\mathcal S_r$: as soon as at least $N$ $s_w$ only occur in some $\mathcal S_i$, then most of the $s_w$ must only occur in that $\mathcal S_i$. We will then see that this is absurd.

Let's now show that there doesn't exist any $\mathcal S_i$ satisfying (\ref{eq:enc_size}). Suppose that there does exist such an $\mathcal S_i$.


For any $w$ of length $h$, let $\structarbre_w$ be the set of $s_{w'}$ such that $w$ is a prefix of $w'$, together with a copy of \loz{p}{l} joining every $s_{w'}$ and $s_{w''}$ where $w''=w'0$ or $w''=w'1$. Basically, $\structarbre_w$ corresponds to the subtree rooted in $w$ in the complete binary trees of universe $\{0,1\}^{\leq n}$, where the edges are replaced with copies of \loz{p}{l}. There are $2^h$ such $\structarbre_w$. Let's call them \textbf{$h$-trees}.

Note that if some $h$-tree $\mathcal T_w$ contains
\begin{itemize}
\item an $\mathcal S_i$-inode,
\item an $\mathcal S_i$-onode,
\item no $s_w$ appearing in $t$,
\end{itemize}
then there must exist an $\mathcal S_i$-bridge included in $\mathcal T_w$. Since, as seen earlier, there cannot exist more than $2k+2$ $\mathcal S_i$-bridges and $t$ contains at most $k+1$ $s_w$, this entails that of the $2^h$ disjoint $h$-trees, at most $3k+3$ can contain at the same time some $\mathcal S_i$-inode and some $\mathcal S_i$-onode.

\bigskip

Recall that we supposed in (\ref{eq:enc_size}) that there were at least $N$ $\mathcal S_i$-inodes. Since there are only $2^h-1$ $s_w$ not belonging to any $h$-tree (those for which $|w|<h$), and since each $h$-tree contains $2^{n-h+1}-1$ $s_w$, by choice of $N$ there must exist at least $4k+5$ $h$-trees containing some $\mathcal S_i$-inode.

Similarly, since the second inequality of (\ref{eq:enc_size}) entails that there exist at least $N$ $\mathcal S_i$-onodes, there must exist at least $4k+5$ $h$-trees contining some $\mathcal S_i$-onode.

We've seen that at most $3k+3$ of the $h$-trees can contain at the same time an $\mathcal S_i$-inode and an $\mathcal S_i$-onode. Thus, there must exist at least $k+2$ $h$-trees $\mathcal T$ containing only $\mathcal S_i$-inodes and $s_w$ appearing in $t$; hence there exists some $h$-tree $\mathcal T$ containing only $\mathcal S_i$-inodes. Similarly, there must exist some $h$-tree $\mathcal T'$ containing only $\mathcal S_i$-onodes.

\bigskip

Let's now finish the construction of $\structcounter$ and $\structcounterbis$ so that the existence of $\structarbre$ and $\structarbrebis$ yields at least $2k+3$ $\mathcal S_i$-bridges. For that, we add copies of \loz{p}{l} between the $s_w$, $|w|=n$ in order to make sure that for every $h$-trees $\structarbre_w\neq\structarbre_{w'}$, there are $2k+3$ leaves of $\structarbre_w$ which are adjacent to leaves of $\structarbre_{w'}$. Once we've shown how to do this, we get $2k+3$ $\mathcal S_i$-bridges involving leaves of $\structarbre$ (which are $\mathcal S_i$-inodes) and leaves of $\structarbrebis$ (which are $\mathcal S_i$-onodes).

Consider an $h$-tree $\structarbre_w$. In $\structarbre_w$, there are $2^{n-h-1}$ leaves $s_{w'}$ with $w'$ ending with a $1$. Those are isolated (not taking into account their label) in $\structcounter$. Since there are $2^h-1$ other $h$-trees, as long as $2^{n-h-1}\geq (2k+3)(2^h-1)$ i.e. in particular for
\begin{equation}
  \label{eq:h}
  h:=\lfloor \frac{n-1-\log(2k+3)}{2}\rfloor\,,
\end{equation}
we can arbitrarily put $2k+3$ copies of \loz{p}{l} between the leaves ending with $1$ of every pair of $h$-tree. This leads to an impossibility, which comes from the assumption of the existence of an $\mathcal S_i$ satisfying (\ref{eq:enc_size}).

At this point, $\structcounter$ and $\structcounterbis$ are still unions of concatenations of copies of \loz{p}{l} (indeed, in the previous step, copies of \loz{p}{l} were only added between isolated $s_w$ of $\structcounter$). As promised, we now arbitrarily add copies of \loz{p}{l} in $\structcounter$ and $\structcounterbis$ so that both of them are a concatenation of copies of \loz{p}{l}.

As desired, both structures have degree $5$, and treewidth at most $2$ (indeed, they are series-parallel graphs). Setting $l:=2^\simindexbis$ ensures that \[\structcounter\foeq{\simindexbis}\structcounterbis\,,\] since the Spoiler in the $\simindexbis$-round \EF game has no way of determining in which order the $s_w$ are linked in $\structcounter$ and $\structcounterbis$.

\bigskip

We've seen that for every bag $t\in S$, there are at most $2k+3$ connected components $\mathcal S_1,\cdots,\mathcal S_r$ in $\mathcal S\setminus\{t\}$. Furthermore, no $\mathcal S_i$ can satisfy (\ref{eq:enc_size}).

It is not possible for all the connected components of $\mathcal S\setminus\{t\}$ to have less than $N$ inodes, as long as $n$ is large enough, since \[(2k+3)(N-1)+(k+1)<2^{n+1}-1\,.\] Hence there must exist some $\mathcal S_i$ with at least $N$ $\mathcal S_i$-inodes. Since $\mathcal S_i$ cannot satisfy (\ref{eq:enc_size}), there must exist more than $(2^{n+1}-1)-(k+1)-N$ $\mathcal S_i$-inodes.

Each other connected component $\mathcal S_j,j\neq i$ must then have less than $N$ $\mathcal S_j$-inode.

This unique $\mathcal S_i$ is called the \textbf{large connected component of $\mathcal S\setminus\{t\}$}.

We are now ready to conclude the proof. For that, consider Algorithm~\ref{alg:counterex_tw}.

\begin{algorithm}[!ht]
  \small
  
  \caption{}
  \label{alg:counterex_tw}
  
  \begin{algorithmic}[1]
    \State Arbitrarily pick a bag $t\in S$
    \While{true}
    \State Print $t$
    \State $t\gets$ the neighbor of $t$ in the large connected component of $\mathcal S\setminus\{t\}$.
    \EndWhile
  \end{algorithmic}
\end{algorithm}

Let's look at the infinite sequence output by Algorithm~\ref{alg:counterex_tw}.

$\mathcal S$ being acyclic and finite, some sequence $t_1,t_2,t_1$, with $t_1,t_2\in S$, must occur at some point in the output string.

Let $\mathcal S_1$ (resp. $\mathcal S_2$) be the connected component of $\mathcal S\setminus\{(t_1,t_2)\}$ containing $t_1$ (resp. $t_2$). The apparition of the sequence $t_1,t_2,t_1$ in the output means that
\begin{itemize}
\item $\mathcal S_2$ is the large connected component of $\mathcal S\setminus\{t_1\}$
\item $\mathcal S_1$ is the large connected component of $\mathcal S\setminus\{t_2\}$.
\end{itemize}

In particular, this means that there exist more that $(2^{n+1}-1)-(k+1)-N$ $\mathcal S_1$-inodes, and more than $(2^{n+1}-1)-(k+1)-N$ $\mathcal S_2$-inodes. However, those sets of inodes are disjoint, and the sum of their number exceeds $2^{n+1}-1$ as long as $n$ is chosen large enough wrt. $k$ and $\delta$ (recall that $N=O(2^{\frac{n}{2}})$).

We have thus reached an impossibility, proving that there cannot exist tree-decompositions $\deccounter$ and $\deccounterbis$ in $\TDD{k}{\delta}$ such that \[\deccounter\foeq{\alpha}\deccounterbis\,.\]
This answers Question~\ref{qu:tw} by the negative.

\section{Conclusion}

We have seen that provided that we restrict ourselves to the very natural notion of decompositions of bounded span, there is no hope to define in \FO, as one is able to do in \MSO, path-decompositions (resp. tree-decompositions) of a structure of bounded pathwidth (resp. bounded treewidth).

We have been able to prove the non-existence of \FO-continuous decompositions through arguments of compactness of the decomposition. If one lifts the condition on the span of the decompositions, our methods cease to work, since in the general setting there is no way to bound the diameter of a decomposition. However, we insist on the fact that decompositions in which there is no bound on the span are of no use in the context of first-order logic, since the equality between nodes becomes inexpressible.

Note that we have proven a similar result for structures of bounded pathwidth and path-decompositions, but that we have not eliminated the possibility for tree-decompositions of bounded span to be definable in structures of bounded pathwidth. We leave this question for further research.

\bibliography{biblio}{}
\bibliographystyle{plain}

\end{document}